\documentclass[conference]{IEEEtran}

\usepackage{amsmath}
\usepackage{amssymb}
\usepackage{amsfonts}
\usepackage{graphicx}
\usepackage{epsfig}
\usepackage{subfigure}
\usepackage{psfrag}
\usepackage{cite}
\usepackage{latexsym}
\usepackage{url}
\usepackage{color}
\usepackage{multirow}

\PassOptionsToPackage{bookmarks={false}}{hyperref}

\begin{document}

\title{Proactive Eavesdropping via Jamming over HARQ-Based Communications}

\author{\IEEEauthorblockN{Jie~Xu\IEEEauthorrefmark{1},
Kai Li\IEEEauthorrefmark{2},
Lingjie~Duan\IEEEauthorrefmark{2}, and
Rui~Zhang\IEEEauthorrefmark{3}}
\IEEEauthorblockA{\IEEEauthorrefmark{1}School of Information Engineering, Guangdong University of Technology}
\IEEEauthorblockA{\IEEEauthorrefmark{2}Engineering Systems and Design Pillar, Singapore University of Technology and Design}
\IEEEauthorblockA{\IEEEauthorrefmark{3}Department of Electrical \& Computer Engineering, National University of Singapore}
Email: jiexu@gdut.edu.cn, kai\_li@sutd.edu.sg, lingjie\_duan@sutd.edu.sg, elezhang@nus.edu.sg
}

\maketitle
\vspace{-1em}
\begin{abstract}
This paper studies the wireless surveillance of a hybrid automatic repeat request (HARQ) based suspicious communication link over Rayleigh fading channels. We propose a proactive eavesdropping approach, where a half-duplex monitor can opportunistically jam the suspicious link to exploit its potential retransmissions for overhearing more efficiently. In particular, we consider that the suspicious link uses at most two HARQ rounds for transmitting the same data packet, and we focus on two cases without and with HARQ combining at the monitor receiver. In both cases, we aim to maximize the successful eavesdropping probability at the monitor, by adaptively allocating the jamming power in the first HARQ round according to fading channel conditions, subject to an average jamming power constraint. For both cases, we show that the optimal jamming power allocation follows a threshold-based policy, and the monitor jams with constant power when the eavesdropping channel gain is less than the threshold. Numerical results show that the proposed proactive eavesdropping scheme achieves higher successful eavesdropping probability than the conventional passive eavesdropping, and HARQ combining can help further improve the eavesdropping performance.
\end{abstract}

\IEEEpeerreviewmaketitle
\setlength\abovedisplayskip{1pt}
\setlength\belowdisplayskip{1pt}

\newtheorem{proposition}{\underline{Proposition}}[section]
\newtheorem{lemma}{\underline{Lemma}}[section]

\section{Introduction}
\label{intro}

Recent advances in user-controlled wireless networks and devices such as ad hoc networks and drones have posed new threats to public security, since they can be misused to facilitate or commit crimes and terror attacks. In order to prevent or defend against such misuse, there is a growing need for authorized parties to legitimately monitor and eavesdrop suspicious communication links. 
In this case, different from conventional wireless security that assumes communication links are rightful and aims to maximize the secrecy rate against illegal eavesdropping~\cite{Zou2016}, we consider a new wireless surveillance paradigm that focuses on legitimately eavesdropping suspicious wireless communication links~\cite{Xu2017Surveillance,xu2017proactive,xu2016proactive,zeng2016wireless,Ma2017,Tran2016,Zhong2017}.

Passive eavesdropping is a commonly adopted wireless surveillance approach, which, however, is unable to overhear the suspicious communications clearly once the legitimate monitors (eavesdroppers) are far away from suspicious transmitters (STs), due to the severe path-loss and channel fading of the eavesdropping link. To cope with this issue, proactive eavesdropping via jamming has been proposed in~\cite{xu2017proactive,xu2016proactive,zeng2016wireless}, where a full-duplex monitor sends jamming signals to interfere with the suspicious receiver (SR) to moderate the suspicious link transmission parameters (such as power and/or rate), for facilitating the simultaneous eavesdropping. By exploiting the channel fluctuations over time, the monitor can adaptively adjust its jamming power based on instantaneous channel conditions, for improving the eavesdropping performance \cite{xu2017proactive}. It is worth noting that the above works largely focus on maximizing the eavesdropping capacity of the monitor assisted with jamming, in which the same packet is transmitted once in the suspicious communication link. In practice, however, most wireless communication systems are operated based on hybrid automatic repeat request (HARQ) protocols to ensure reliable communications, where the transmitter may retransmit the same packet if the receiver fails to decode \cite{LTE,Cheng2006}. Furthermore, the existing works assume full-duplex monitors with simultaneous jamming and eavesdropping, but the performance is practically limited by the self-interference from the jamming to the eavesdropping antennas at the monitor~\cite{Sabharwal2014}.

To overcome these limitations, this paper studies the wireless surveillance of an HARQ-based suspicious communication link via a practical half-duplex legitimate monitor over Rayleigh fading channels. We consider that the suspicious communication link implements the HARQ protocol as follows. Initially, the ST transmits a coded packet to the SR; depending on whether the SR decodes it successfully or not, it replies an acknowledgement (ACK) or negative acknowledgement (NACK); upon receiving a NACK, the ST retransmits the same coded packet again. This operation is repeated until either an ACK is received or the number of retransmissions exceeds a maximum threshold. Under this setup, we propose a proactive eavesdropping approach, where the monitor opportunistically jams the suspicious link to exploit its potential retransmissions for overhearing more efficiently.

In particular, we assume the number of HARQ rounds for the suspicious communication to be two with at most one retransmission for the same packet.\footnote{If more than one retransmissions are allowed, our eavesdropping performance is expected to be further improved, as there will be more chances for the monitor to jam and eavesdrop the suspicious link retransmissions.} In this case, the monitor must work in the eavesdropping mode without any jamming in the second HARQ round, since the half-duplex monitor cannot jam and eavesdrop the suspicious link at the same time, and there will be no more retransmissions of the same packet that can be eavesdropped. We focus on two cases without and with HARQ combining at the monitor receiver, which decodes each retransmitted packet independently, or combines all previously received copies of the same packet to decode with maximum ratio combining (MRC), respectively.\footnote{With the same packet retransmitted, the considered HARQ protocol with MRC is referred to as ``chase combining (CC)'' in practice \cite{Cheng2006}. There is another HARQ protocol with combining, namely ``incremental redundancy (IR)'', where if a NACK is received, the ST transmits additional coded bits, instead of retransmitting the same packet. Our results in this paper can also be extended to the case of IR, which is left for future work.} In both cases, we aim to maximize the successful eavesdropping probability at the monitor, by adaptively allocating the jamming power in the first HARQ round based on fading channel conditions, subject to an average jamming power constraint. For both cases, we show that the optimal jamming power allocation follows a threshold-based policy, in which the monitor jams with constant power when the eavesdropping channel gain is less than a threshold; otherwise, the jamming power is zero and the monitor eavesdrops in the first round. Specifically, we find that with HARQ combining, the threshold is generally smaller than the case without HARQ combining; therefore, in this case the monitor prefers jamming only when the eavesdropping channel gain is even weaker. Finally, numerical results show that the proposed proactive eavesdropping schemes achieve higher successful eavesdropping probability than the conventional passive eavesdropping, and HARQ combining can help further improve the eavesdropping performance.

\begin{figure}[t]
\centering
\includegraphics[width=0.4\textwidth]{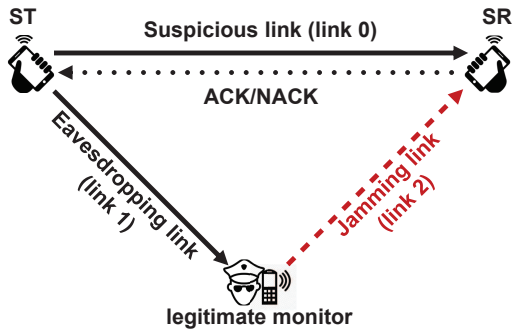}
\caption{Wireless surveillance of an HARQ-based suspicious communication link, where the legitimate monitor opportunistically jams the suspicious link to increase the chance of its retransmission for proactive eavesdropping.}
\label{fig_app}\vspace{-1em}
\end{figure}



\section{System Model}
\label{system}
As shown in Fig.~\ref{fig_app}, a half-duplex legitimate monitor aims to overhear a suspicious communication link from an ST to an SR, which employs the HARQ protocol with two transmission rounds at most for the same packet. In this case, the monitor must work in the eavesdropping mode without any jamming in the second HARQ round, and can jam or eavesdrop (but not at the same time) in the first round. Intuitively, jamming the SR in the first round can increase the probability of packet retransmission in the second round of the suspicious link, but lose the chance of eavesdropping it in the first round.

We consider a quasi-static channel model, where wireless channels remain unchanged at each transmission round of one packet, but can change independently over different rounds and for different packets. Let $g^{t}_{i}$ denote the channel power gain of link $i$ at the $t$-th transmission round for one packet, where $i = 0$, $1$, and $2$ represent the suspicious link, the eavesdropping link, and the jamming link, respectively. Here, $t = \text{I}$ and $t = \text{II}$ denote the initial transmission and the retransmission rounds of the same packet, respectively. Rayleigh fading is considered, where for any $t \in \{\text{I},\text{II}\}$, $g^{t}_{i}$ follows an exponential distribution with the rate parameter $\lambda_{i}$ (or the mean value $1/\lambda_{i}$), $i\in\{0,1,2\}$. We consider that at each transmission round $t$, the channel state information (CSI) $g^{t}_{0}$ of the suspicious link is only available at the SR. Therefore, the ST adopts a fixed transmit power $P_{0}$ and a fixed data rate $R$ (in bps/Hz) to deliver different packets over time. It is also assumed that at the beginning of round $t = \text{I}$, the monitor perfectly knows the CSI $g^{ \text{I}}_{1}$ of the eavesdropping link via efficient channel estimation based on the received pilot signal from the ST, and it also knows the channel distribution information (CDI) of all the three links (i.e., the values of $\lambda_{i}$'s) based on long-term observation.

Based on the CSI $g^{\text{I}}_{1}$ of the eavesdropping link, in round $t = \text{I}$ the monitor can adaptively adjust its power for jamming or just eavesdrop without jamming to maximize the surveillance performance over the HARQ-based suspicious communication. Let $Q(g^{\text{I}}_{1}) \ge 0$ denote the jamming power in round $t = \text{I}$ based on the exactly known $g^{\text{I}}_{1}$, where $Q(g^{\text{I}}_{1}) = 0$ tells that the monitor eavesdrops without jamming in the first round. Our objective is to optimize the jamming power $Q(g^{\text{I}}_{1})$'s according to the eavesdropping channel condition $g_1^{\text{I}}$'s over time to maximize the successful eavesdropping probability (to be defined later), subject to an average jamming power budget. We address this problem in the following by considering two cases without and with HARQ combining at the monitor receiver, respectively.

\section{Optimal Jamming without HARQ  Combining}
\label{sec:without:combining}

In this section, we consider that if the monitor fails to decode the packet in the initial transmission round $t = \text{I}$, it will discard the packet in this round; then in any retransmission round $t = \text{II}$, it will only use the retransmitted packet for decoding, for ease of implementation. In the following, we first derive the successful eavesdropping probability under given jamming power $Q(g^{\text{I}}_{1})$ for any given $g^{\text{I}}_{1}$. Then, we decide the jamming power allocation to maximize the successful eavesdropping probability by considering all possible values of $g^{\text{I}}_{1}$'s over time given the average jamming power budget.

\subsection{Successful Eavesdropping Probability Under Given $g^{\text{I}}_{1}$ and $Q(g^{\text{I}}_{1})$}

In this subsection, we obtain the successful eavesdropping probability under any given $g^{\text{I}}_{1}$ and $Q(g^{\text{I}}_{1})$, denoted as $\mathcal{P}_{\text{eav}} (g^{\text{I}}_{1},Q(g^{\text{I}}_{1}))$. In particular, we have either $Q(g^{\text{I}}_{1}) > 0$ (in jamming mode) or $Q(g^{\text{I}}_{1}) = 0$ (in eavesdropping mode) in round $t = \text{I}$.

\subsubsection{Jamming with $Q(g^{\text{I}}_{1}) > 0$}

In this case, the monitor can only overhear the retransmission in round $t = \text{II}$, provided that the suspicious transmission in round $t = \text{I}$ fails. First, we obtain the probability of suspicious retransmission, which is equivalent to the outage probability at the SR after the initial transmission of the ST in round $t = \text{I}$. In this round, the received signal-to-interference-and-noise ratio (SINR) at the SR is
\begin{equation}
\gamma_{0}^{\text{I}}(Q(g^{\text{I}}_{1})) = \frac{g^{\text{I}}_{0}P_{0}}{g^{\text{I}}_{2}Q(g^{\text{I}}_{1}) + \sigma^{2}},
\end{equation}
where $\sigma^{2}$ denotes the noise power at the receiver. Note that $g^{\text{I}}_{0}$ and $g^{\text{I}}_{2}$ are independent exponentially distributed variables with rate parameters $\lambda_{0}$ and $\lambda_{2}$, respectively. By letting \begin{align}\label{eqn:gamma}
\bar{\gamma}=2^R-1
 \end{align}
denote the minimum received signal-to-noise ratio (SNR) or SINR requirement for successfully decoding the packet by assuming the optimal Gaussian signalling employed, we have the probability of suspicious retransmission as \cite{xu2016proactive}
\begin{align}
&p^{\text{I-out}}_{0}(Q(g^{\text{I}}_{1})) = \mathbb{P}(\gamma_{0}^{\text{I}}(Q(g^{\text{I}}_{1})) < \bar{\gamma}) \nonumber \\
= &1 - \frac{\lambda_{2}/(\bar{\gamma} Q(g^{\text{I}}_{1}))}{\lambda_{0}/P_0 + \lambda_{2}/(\bar{\gamma} Q(g^{\text{I}}_{1}))} e^{-\lambda_{0}\sigma^{2}\bar{\gamma}/P_0}.
\label{eq_jamming_p01}
\end{align}

Next, we obtain the conditional successful eavesdropping probability of the monitor in round $t = \text{II}$, denoted by $p^{\text{II-suc}}_{1}$. In this round, the received SNR at the monitor is given as $\gamma_{1}^{\text{II}} = {g^{\text{II}}_{1}P_{0}}/{\sigma^{2}}$. As $g^{\text{II}}_{1}$ is exponentially distributed with the rate parameter $\lambda_{1}$, we have
\begin{align}
p^{\text{II-suc}}_{1} =& \mathbb{P}(\gamma_{1}^{\text{II}} \ge \bar{\gamma}) = e^{-\lambda_{1}\sigma^{2}\bar{\gamma}/P_0}. \label{eq_jamming_p12out}
\end{align}
By combining the probability $p^{\text{I-out}}_{0}(Q(g^{\text{I}}_{1}))$  of suspicious retransmission in (\ref{eq_jamming_p01}) and the conditional successful eavesdropping probability $p^{\text{II-suc}}_{1}$ in (\ref{eq_jamming_p12out}), we have the successful eavesdropping probability under given $g^{\text{I}}_{1}$ and $Q(g^{\text{I}}_{1}) > 0$ as
\begin{align}
&\mathcal{P}_{\text{eav}}(g^{\text{I}}_{1},Q(g^{\text{I}}_{1}))  = p^{\text{II-suc}}_{1} p^{\text{I-out}}_{0}(Q(g^{\text{I}}_{1})) \nonumber\\
 =~& e^{-\lambda_{1}\sigma^{2}\bar{\gamma}/P_0} - \frac{\lambda_{2}P_0}{\lambda_{0}\bar{\gamma} Q(g^{\text{I}}_{1}) + \lambda_{2}P_0} e^{-(\lambda_{0}+\lambda_{1})\sigma^{2}\bar{\gamma}/P_0} \nonumber\\
 \triangleq ~ & \Phi(Q(g^{\text{I}}_{1})).\label{eqn:eav:succ}
\end{align}
Note that the function $\Phi(Q(g^{\text{I}}_{1}))$ is independent of $g^{\text{I}}_{1}$, and is monotonically increasing and concave with respect to the jamming power $Q(g^{\text{I}}_{1}) \ge 0$.

\subsubsection{Eavesdropping with $Q(g^{\text{I}}_{1}) = 0$}\label{Sec:Eav}

In this case, the received SNR at the monitor is $\gamma_{1}^{\text{I}} = {g^{\text{I}}_{1}P_{0}}/{\sigma^{2}}$.
If $\gamma_{1}^{\text{I}}$ is no smaller than the minimum SNR requirement $\bar\gamma$ in \eqref{eqn:gamma} (i.e, $\gamma_{1}^{\text{I}} \ge \bar\gamma$), or equivalently $g^{\text{I}}_{1} \ge \bar g \triangleq {\bar{\gamma}\sigma^{2}}/{P_0}$, the monitor can successfully decode the ST's transmitted packet in this round, no matter whether the retransmission of the suspicious link occurs or not. Here, we have the successful eavesdropping probability under given $g^{\text{I}}_{1} \ge \bar g$ and $Q(g^{\text{I}}_{1}) = 0$ as
\begin{align}\label{eqn:2}
\mathcal{P}_{\text{eav}} (g^{\text{I}}_{1},Q(g^{\text{I}}_{1}))  =  1.
\end{align}

On the other hand, if $\gamma_{1}^{\text{I}} < \bar\gamma$ or equivalently $g^{\text{I}}_{1} < \bar g$, the monitor cannot decode the packet in round $t = \text{I}$. Therefore, it can only overhear the retransmission in round $t = \text{II}$, under the condition that the suspicious transmission in round $t = \text{I}$ fails. The successful eavesdropping probability in this case can be similarly obtained as that in (\ref{eqn:eav:succ}). By replacing $Q(g^{\text{I}}_{1}) > 0$ in (\ref{eqn:eav:succ}) as $Q(g^{\text{I}}_{1}) = 0$, we have the successful eavesdropping probability under given $g^{\text{I}}_{1} < \bar g$ and $Q(g^{\text{I}}_{1}) = 0$ as
\begin{align}\label{eqn:2:2}
\mathcal{P}_{\text{eav}} (g^{\text{I}}_{1},Q(g^{\text{I}}_{1})) = \Phi(0),
\end{align}
where $\Phi(\cdot)$ is given in (\ref{eqn:eav:succ}).

By combining $\mathcal{P}_{\text{eav}} (g^{\text{I}}_{1},Q(g^{\text{I}}_{1}))$ in (\ref{eqn:eav:succ}), (\ref{eqn:2}), and (\ref{eqn:2:2}), it follows that the successful eavesdropping probability under any given $g^{\text{I}}_{1}$ and $Q(g^{\text{I}}_{1})$ is obtained as
\begin{align}\label{eqn:2:2:ALL}
&\mathcal{P}_{\text{eav}} (g^{\text{I}}_{1},Q(g^{\text{I}}_{1})) =
\left\{
\begin{array}{ll}
\Phi(Q(g^{\text{I}}_{1})), & {\text{if}}~ Q(g^{\text{I}}_{1}) > 0,\\
1, & {\text{if}}~ Q(g^{\text{I}}_{1}) = 0~{\text{and}}~g^{\text{I}}_{1} \ge \bar g,
\\
\Phi(0), & {\text{if}}~ Q(g^{\text{I}}_{1}) = 0~{\text{and}}~g^{\text{I}}_{1} < \bar g.
\end{array}
\right.
\end{align}
\vspace{-1em}
\subsection{Successful Eavesdropping Probability Maximization}

Our objective is to maximize the successful eavesdropping probability over all possible $g_1^{\text{I}}$'s, subject to the average jamming power budget at the monitor, denoted by $Q_{\text{ave}} > 0$. Towards this end, we adaptively allocate the jamming power $Q(g^{\text{I}}_{1})$'s based on the exact CSI observation of $g_1^{\text{I}}$'s for different packets. The optimization problem for the monitor is formulated as
\begin{align}
\text{(P1)}:\max_{\{Q(g^{\text{I}}_{1}) \ge 0\}}~& \mathbb{E}_{g^{\text{I}}_{1}}\left(\mathcal{P}_{\text{eav}} (g^{\text{I}}_{1},Q(g^{\text{I}}_{1}))\right) \nonumber\\
\text{s.t.}~~~~~&\mathbb{E}_{g^{\text{I}}_{1}}\left(Q(g^{\text{I}}_{1})\right) \le Q_{\text{ave}},
\end{align}
where $\mathbb{E}_{g^{\text{I}}_{1}}(\cdot)$ denotes the expectation operation over $g^{\text{I}}_{1}$. We have the following proposition.
\begin{proposition}\label{propo:1}
The optimal jamming power solution to problem (P1) is given as
\begin{align}
Q^*(g^{\text{I}}_{1}) = \left\{
\begin{array}{ll}
0, &\forall g^{\text{I}}_{1} \ge \bar g\\
\frac{Q_{\text{ave}}}{1 - e^{-\lambda_{1}\sigma^{2}\bar{\gamma}/P_0}},~&\forall g^{\text{I}}_{1} < \bar g
\end{array}
\right..
\end{align}
\end{proposition}
\begin{IEEEproof}
Under any packet transmission with $g^{\text{I}}_{1} \ge \bar g$, it is evident that setting the jamming power as $Q(g^{\text{I}}_{1}) = 0$ achieves $\mathcal{P}_{\text{eav}} (g^{\text{I}}_{1},Q(g^{\text{I}}_{1})) = 1$ in \eqref{eqn:2}, while setting $Q(g^{\text{I}}_{1}) > 0$  achieves $\mathcal{P}_{\text{eav}} (g^{\text{I}}_{1},Q(g^{\text{I}}_{1})) < 1$ in \eqref{eqn:eav:succ}. Therefore, we have $Q^*(g^{\text{I}}_{1}) = 0$ for any $g^{\text{I}}_{1} \ge \bar g$.

Next, consider another packet transmission with $g^{\text{I}}_{1} < \bar g$. In this case, by combining \eqref{eqn:eav:succ} and (\ref{eqn:2:2}), it is evident that $\mathcal{P}_{\text{eav}}(g^{\text{I}}_{1},Q(g^{\text{I}}_{1}))$ is a monotonically increasing and concave function with respect to the jamming power $Q(g^{\text{I}}_{1}) \ge 0$, but irrespective of $g^{\text{I}}_{1}$. Based on the Jensen's inequality, it is optimal to set the jamming power to be identical, i.e., $Q^*(g^{\text{I}}_{1}) = Q^*, \forall g^{\text{I}}_{1} < \bar g$. Furthermore, note that the optimality of (P1) is achieved when the average jamming power constraint is met with strict equality. Notice that $\mathbb{P}(g^{\text{I}}_{1} < \bar g) = 1 - e^{-\lambda_{1}\sigma^{2}\bar{\gamma}/P_0}$. Then it follows that $(1 - e^{-\lambda_{1}\sigma^{2}\bar{\gamma}/P_0}) \cdot Q^* = Q_{\text{ave}}$. Therefore, we have $Q^*(g^{\text{I}}_{1}) = Q^* = \frac{Q_{\text{ave}}}{1 - e^{-\lambda_{1}\sigma^{2}\bar{\gamma}/P_0}}$ for any $g^{\text{I}}_{1} < \bar g$.

By combining the above two scenarios, this proposition is verified.
\end{IEEEproof}

From Proposition \ref{propo:1}, it is observed that the optimal jamming power allocation follows a threshold-based policy. When the eavesdropping channel gain $g^{\text{I}}_{1}$ is larger than or equal to the threshold $\bar g$, the monitor does not jam as it can successfully eavesdrop; otherwise, when $g^{\text{I}}_{1}$ is smaller than $\bar g$, it is optimal for the monitor to employ constant-power jamming to maximize the probability of retransmission and hence the successful eavesdropping probability.

\section{Optimal Jamming with HARQ Combining}
\label{extension}

In this section, we consider that if the SR or the monitor fails to decode the packet eavesdropped in the initial transmission ($t = \text{I}$), it will use it to combine with the retransmitted packet in the second round ($t = \text{II}$) via the MRC technique.

\subsection{Successful Eavesdropping Probability Under Given $g^{\text{I}}_{1}$ and $Q(g^{\text{I}}_{1})$}
In the following, we obtain the successful eavesdropping probability under any given $g^{\text{I}}_{1}$ and $Q(g^{\text{I}}_{1})$, denoted as $\hat{\mathcal{P}}_{\text{eav}} (g^{\text{I}}_{1},Q(g^{\text{I}}_{1}))$.

\subsubsection{Jamming  with $Q(g^{\text{I}}_{1}) > 0$}

In this case, eavesdropping is not feasible at the monitor in round $t = \text{I}$, and hence, no MRC is implementable. Therefore, the successful eavesdropping probability under given $g^{\text{I}}_{1}$ and $Q(g^{\text{I}}_{1}) > 0$ is same as that in (\ref{eqn:eav:succ}) without HARQ combining, i.e., $\hat{\mathcal{P}}_{\text{eav}} (g^{\text{I}}_{1},Q(g^{\text{I}}_{1})) = \Phi(Q(g^{\text{I}}_{1}))$.

\subsubsection{Eavesdropping with $Q(g^{\text{I}}_{1}) = 0$}

When $g^{\text{I}}_{1} \ge \bar g$, the eavesdropping is always successful in round $t = \text{I}$. Therefore, we have the successful eavesdropping probability under given $g^{\text{I}}_{1} \ge \bar g$ and $Q(g^{\text{I}}_{1}) = 0$ as $\hat{\mathcal{P}}_{\text{eav}} (g^{\text{I}}_{1},Q(g^{\text{I}}_{1})) =1$.

When $g^{\text{I}}_{1} < \bar g$, the eavesdropping is not successful at round $t=\text{I}$. Nevertheless, if retransmission occurs, the monitor can implement MRC to combine the two copies of the same packet that are received in the two rounds, respectively. First, note that the probability of retransmission can be similarly obtained as that in (\ref{eq_jamming_p01}) by calculating the outage probability of suspicious transmission. As no jamming is employed here, we have the probability of retransmission as
\begin{align}\label{eqn:re:tran:0}
p^{\text{I-out}}_{0}(0) = 1-e^{-\lambda_{0}\sigma^{2}\bar{\gamma}/P_0},
\end{align}
with $p^{\text{I-out}}_{0}(Q(g^{\text{I}}_{1}))$ given in (\ref{eq_jamming_p01}).

Next, we derive the conditional successful eavesdropping probability under given $g^{\text{I}}_{1}$ when retransmission occurs, denoted as $\hat{p}^{{\text{I+II-suc}}}_1(g^{\text{I}}_{1})$. With MRC, the received SNR at the monitor is given as
\begin{align}
\gamma_{1}^{\text{I}} + \gamma_{1}^{\text{II}} = \frac{(g^{\text{I}}_{1}+g^{\text{II}}_{1})P_0}{\sigma^2}.
\end{align}
As $g^{\text{II}}_{1}$ is exponentially distributed with rate parameter $\lambda_{1}$, we have
\begin{align}
&\hat{p}^{{\text{I+II-suc}}}_1(g^{\text{I}}_{1}) = \mathbb{P}\left(\gamma_{1}^{\text{I}} + \gamma_{1}^{\text{II}} \ge \bar{\gamma}\right) = \mathbb{P}\left(g^{\text{II}}_{1} \ge \frac{\sigma^{2}\bar{\gamma}}{P_0}  - g^{\text{I}}_{1} \right) \nonumber \\
&=e^{-\frac{\bar{\gamma}\lambda_{1}\sigma^{2}}{P_0} + g^{\text{I}}_{1}\lambda_{1}}.
\label{eq_mrc_p1out}
\end{align}
Then, by taking into account the
probability of retransmission $p^{\text{I-out}}_{0}(0)$ in (\ref{eqn:re:tran:0}), we have the successful eavesdropping probability under given $g^{\text{I}}_{1} < \bar g$ and $Q(g^{\text{I}}_{1}) = 0$ as
\begin{align}
&\hat{\mathcal{P}}_{\text{eav}}(g^{\text{I}}_{1},Q(g^{\text{I}}_{1})) = \hat{p}^{{\text{I+II-suc}}}_1(g^{\text{I}}_{1})\cdot p^{\text{I-out}}_{0}(0) \nonumber\\
= & e^{-\bar{\gamma}\lambda_{1}\sigma^{2}_{1}/P_0 + g^{\text{I}}_{1}\lambda_{1}}(1-e^{-\lambda_{0}\sigma^{2}\bar{\gamma}/P_0}) = e^{g^{\text{I}}_{1}\lambda_{1}}\Phi(0).
\end{align}

By combining the above cases, it follows that with HARQ combining, the successful eavesdropping probability under any given $g^{\text{I}}_{1}$ and $Q(g^{\text{I}}_{1})$ is
\begin{align}\label{eqn:2:2:2}
&\hat{\mathcal{P}}_{\text{eav}} (g^{\text{I}}_{1},Q(g^{\text{I}}_{1})) =
\left\{
\begin{array}{ll}
\Phi(Q(g^{\text{I}}_{1})), & {\text{if}}~ Q(g^{\text{I}}_{1}) > 0,\\
1, & {\text{if}}~ Q(g^{\text{I}}_{1}) = 0~{\text{and}}~g^{\text{I}}_{1} \ge \bar g,
\\
e^{g^{\text{I}}_{1}\lambda_{1}}\Phi(0), & {\text{if}}~ Q(g^{\text{I}}_{1}) = 0~{\text{and}}~g^{\text{I}}_{1} < \bar g.
\end{array}
\right.
\end{align}
By comparing \eqref{eqn:2:2:2} with \eqref{eqn:2:2:ALL}, it is evident that their only difference lies in the case when the monitor chooses to overhear in round $t = \text{I}$ (i.e., $Q(g^{\text{I}}_{1}) = 0$) but the eavesdropping in round $t=\text{I}$ is not successful (i.e., $g^{\text{I}}_{1} < \bar g$). Thanks to the MRC, the successful eavesdropping probability increases by a factor of $e^{g^{\text{I}}_{1}\lambda_{1}} > 1$ in (\ref{eqn:2:2:2}).

\subsection{Successful Eavesdropping Probability Maximization}

With HARQ combining, the successful eavesdropping probability maximization problem is formulated as
\begin{align}
\text{(P2)}:\max_{\{Q(g^{\text{I}}_{1}) \ge 0\}}~& \mathbb{E}_{g^{\text{I}}_{1}}\left(\hat{\mathcal{P}}_{\text{eav}} (g^{\text{I}}_{1},Q(g^{\text{I}}_{1}))\right) \nonumber\\
\text{s.t.}~&\mathbb{E}_{g^{\text{I}}_{1}}\left(Q(g^{\text{I}}_{1})\right) \le Q_{\text{ave}}.
\end{align}
Note that problem (P2) is more challenging to solve than (P1). This is due to the fact that under any given $g^{\text{I}}_{1} < \bar g$, the function $\hat{\mathcal{P}}_{\text{eav}} (g^{\text{I}}_{1},Q(g^{\text{I}}_{1}))$ is non-convex as it is discontinuous for $Q(g^{\text{I}}_{1}) = 0$. Despite this fact, we have the following proposition.
\begin{proposition}\label{proposition:4.1}
The optimal jamming power allocation $\{Q^\star(g^{\text{I}}_{1})\}$ to problem (P2) follows a threshold-based policy, given by
\begin{align}\label{Q:opt}
Q^\star(g^{\text{I}}_{1}) =
&\left\{
\begin{array}{ll}
\sqrt{\frac{\lambda_{2}P_0}{\lambda_{0}\bar{\gamma}\mu^\star}e^{-(\lambda_{0}\sigma^{2}_{0}+\lambda_{1}\sigma^{2}_{1})\bar{\gamma}/P_0}} - \frac{\lambda_{2}P_0}{\lambda_{0}\bar{\gamma}}, & \forall g_{1}^{\text{I}} \le \bar g^\star\\
0, & \forall g_{1}^{\text{I}} > \bar g^\star,
\end{array}
\right.
\end{align}
with the threshold $\bar g^\star$ given as\begin{small}
\begin{align}\label{18}
\bar g^\star = \min \bigg(& \bar g,\frac{1}{\lambda_{1}}\ln \bigg(1 + \frac{\bigg(\sqrt{\frac{\mu^\star\lambda_2 P_0}{\lambda_{0}\bar{\gamma}}} - \sqrt{e^{-\frac{(\lambda_{0}\sigma^{2}+\lambda_{1}\sigma^{2})\bar{\gamma}}{P_0}}}\bigg)^2}{e^{-\frac{\lambda_{1}\sigma^{2}\bar{\gamma}}{P_0}} - e^{-\frac{(\lambda_{0}+\lambda_{1})\sigma^{2}\bar{\gamma}}{P_0}}}\bigg)\bigg).
\end{align}\end{small}Here, $\mu^\star$ satisfying
\begin{align}\label{eqn:mu:star}
0\le \mu^\star \le \frac{\lambda_{0}\bar{\gamma}}{\lambda_{2}P_0}e^{-(\lambda_{0}\sigma^{2}_{0}+\lambda_{1}\sigma^{2}_{1})\bar{\gamma}/P_0}
\end{align}
corresponds to a parameter such that $\mathbb{E}(Q^\star(g^{\text{I}}_{1})) = Q_{\text{ave}}$.
\end{proposition}
\begin{IEEEproof}
See Appendix \ref{appendix_A}.
\end{IEEEproof}

\begin{figure}[t]
\centering
\includegraphics[width=3.3in]{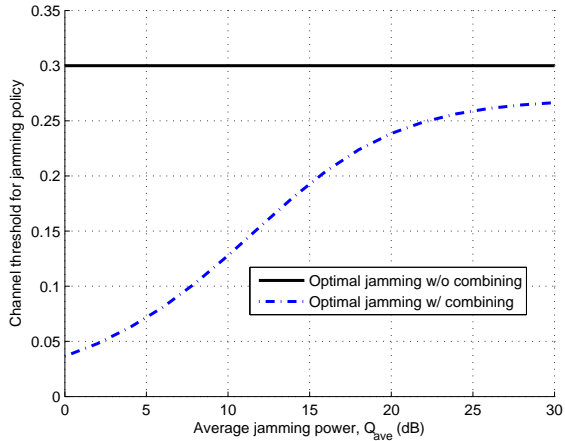}
\caption{The thresholds for jamming policy in the two HARQ cases without versus with combining.}
\label{fig_Threshold}
\vspace{-0em}
\end{figure}

By comparing Proposition \ref{proposition:4.1} with Proposition \ref{propo:1}, it is observed from \eqref{18} that the threshold $\bar g^\star$ for (P2) is smaller than $\bar g$ for (P1). This shows that with the HARQ combining, the monitor may not jam even when it cannot successfully eavesdrop in the initial round. For illustration, we provide a numerical example  in Fig. \ref{fig_Threshold} to show the thresholds for (P1) and (P2), for which the parameters are set as in Section \ref{simulation} later. It is observed that with the jamming power $Q_{\text{ave}}$ increasing, the threshold $\bar g^\star$ for the case with HARQ combining increases monotonically. This indicates that with more jamming power, the monitor should distribute its jamming power over more packets when eavesdropping fails in the initial round.

\section{Numerical Results}
\label{simulation}
In this section, we provide numerical results to evaluate the performance of our proposed proactive eavesdropping schemes as compared to the conventional passive eavesdropping without jamming, i.e., $Q(g_1^{\text{I}}) = 0, \forall g_1^{\text{I}}$. In the simulation, we normalize the noise powers at the SR and the monitor as $\sigma^2 = 1$, set the average channel power gain of the Rayleigh fading suspicious, eavesdropping, and jamming links as $1/\lambda_0 = 1, 1/\lambda_1 = 0.2$, and $1/\lambda_2 = 0.2$, respectively. Furthermore, we fix the transmit power by the ST as $P_0 = 10$ dB.

\begin{figure}[t]
\centering
\includegraphics[width=3.3in]{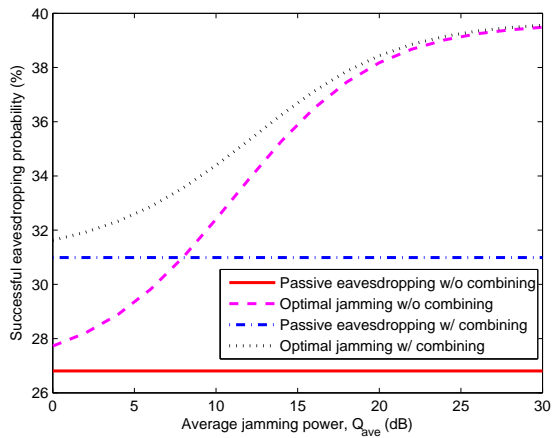}
\caption{The successful eavesdropping probability versus the average jamming power $Q_{\text{ave}}$,  where $R = 2$ bps/Hz.}
\label{fig_eavesdropProb}\vspace{-0em}
\end{figure}

Fig.~\ref{fig_eavesdropProb} shows the successful eavesdropping probability at the monitor versus the average jamming power $Q_{\text{ave}}$, where the communication rate at the suspicious link is $R = 2$ bps/Hz. It is observed that no matter without or with HARQ jamming, the proposed proactive eavesdropping achieves much higher successful eavesdropping probability than the respective passive eavesdropping case. It is also observed that for both passive and proactive eavesdropping, HARQ combining generally helps further improve the eavesdropping performance. Nevertheless, when $Q_{\text{ave}}$ becomes large, the proactive eavesdropping with HARQ combining is observed to achieve a similar eavesdropping performance as that without HARQ combining. This can be explained based on Fig. \ref{fig_Threshold}, where as $Q_{\text{ave}}$ becomes large, the threshold $\bar g^\star$ with HARQ combining increases towards $\bar g$, the threshold for the case without combing. In this case, the monitor needs to jam over more packets when the eavesdropping fails in the initial round, over which the HARQ combining is not applicable, thus making the two schemes perform similar.

%
%
%
%
%
%
%
%


\begin{figure}[htb]
\centering
\includegraphics[width=3.3in]{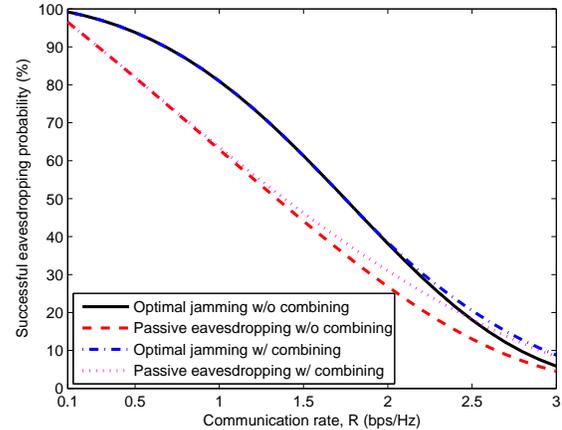}
\caption{The successful eavesdropping probability versus the communication rate $R$ at the suspicious communication, where $Q_{\text{ave}} = 20$dB.}
\label{fig_rateProb}\vspace{-0em}
\end{figure}

Fig.~\ref{fig_rateProb} shows the successful eavesdropping probability versus the suspicious communication rate $R$, where the average jamming power is $Q_{\text{ave}} = 20$ dB. It is observed that at small $R$ values, passive (proactive) eavesdropping with HARQ combining achieves similar eavesdropping performance as that without HARQ combining. This is due to the fact that with small $R$, the eavesdropping is likely to be successful in the initial round, and thus no HARQ combining is required. By contrast, at large $R$ values, eavesdropping with HARQ combining is observed to perform better than that without HARQ combining, as retransmission occurs with a higher probability.

\section{Conclusion}
\label{cond}
This paper presented a new wireless surveillance scenario over an HARQ-based suspicious communication link via a half-duplex monitor. We proposed a proactive eavesdropping via jamming based approach, where the monitor jams opportunistically to improve the surveillance performance via exploiting the potential retransmission in the suspicious link. In both cases without and with HARQ combining at the monitor receiver, we showed that the optimal jamming power allocation follows a threshold-based policy, where the monitor jams with a constant power in the initial round when the eavesdropping channel gain is less than a threshold. It is hoped that this paper can provide new insights on exploiting the retransmission in practical HARQ protocols to improve the performance of wireless surveillance.


\appendix
\subsection{Proof of Proposition \ref{proposition:4.1}}
\label{appendix_A}
Though problem (P2) is non-convex in general, it satisfies the so-called time-sharing condition in \cite{StrongDuality1}. Therefore, strong duality or zero duality gap holds between problem (P2) and its dual problem. Therefore, we use the Lagrange duality method to obtain the optimal solution.

Let $\mu \ge 0$ denote the dual variable associated with the average jamming power constraint. Then the partial Lagrangian of (P2) is denoted as
\begin{align}
\mathcal{L}(\{Q(g^{\text{I}}_{1})\},\mu) = &\mathbb{E}_{g^{\text{I}}_{1}} \left(\hat{\mathcal{P}}_{\text{eav}}(g^{\text{I}}_{1},Q(g^{\text{I}}_{1}))\right)  \nonumber\\
&- \mu\left(\mathbb{E}_{g^{\text{I}}_{1}} \left(Q(g^{\text{I}}_{1})\right) - Q_{\text{ave}}\right).
\end{align}
The dual function is given as
\begin{align}\label{eqn:dual:function}
\psi(\mu) = \max_{\{Q(g^{\text{I}}_{1}) \ge 0\}}\mathcal{L}(\{Q(g^{\text{I}}_{1})\},\mu).
\end{align}
The dual problem is
\begin{align}\label{eqn:dual:problem}
\text{(D2)}:\min_{\mu \ge 0}\psi(\mu).
\end{align}
Due to the strong duality between (P2) and (D2), we solve (P2) by first solving problem (\ref{eqn:dual:function}) to obtain the dual function $\psi(\mu)$ under any given $\mu \ge 0$, and then solving (D2) via searching over $\{Q(g^{\text{I}}_{1})\}$ to minimize $g(\mu)$.

\subsubsection{Solving Problem (\ref{eqn:dual:function}) Under Given $\mu \ge 0$}
Problem (\ref{eqn:dual:function}) can be decomposed into various subproblems each for one $g^{\text{I}}_{1}$.
\begin{align}\label{eqn:sub:pro}
\max_{Q(g^{\text{I}}_{1}) \ge 0} \hat{\mathcal{P}}_{\text{eav}}(Q(g^{\text{I}}_{1})) - \mu Q(g^{\text{I}}_{1}).
\end{align}
We solve problem \eqref{eqn:sub:pro} by considering $g^{\text{I}}_{1}\ge \bar g$ and $g^{\text{I}}_{1} < \bar g$, respectively. First, for any $g^{\text{I}}_{1} \ge \bar g$, it is evident that the optimal solution to problem (\ref{eqn:sub:pro}) is given as
\begin{align}\label{eqn:QQQ:1}
\tilde{Q}(g^{\text{I}}_{1}) = 0, \forall g^{\text{I}}_{1}\ge \bar g,
\end{align}
with the achieved optimal value being $1$.

Next, consider problem (\ref{eqn:sub:pro}) for any given $g^{\text{I}}_{1} < \bar g$. In this case, problem (\ref{eqn:sub:pro}) can be solved by considering two regimes with $Q(g^{\text{I}}_{1}) = 0$ and $Q(g^{\text{I}}_{1}) > 0$, respectively.

In the regime with $Q(g^{\text{I}}_{1}) = 0$, the achieved objective value for problem (\ref{eqn:sub:pro}) is a constant $v_0(g^{\text{I}}_{1})  =  e^{g^{\text{I}}_{1}\lambda_{1}}\Phi(0)$.

In the other regime with $Q(g^{\text{I}}_{1}) > 0$, problem (\ref{eqn:sub:pro}) becomes
\begin{align}
\max_{Q(g^{\text{I}}_{1}) > 0} &
\Phi(Q(g^{\text{I}}_{1}))- \mu Q(g^{\text{I}}_{1}).\label{eqn:25}
\end{align}
As $\Phi(Q(g^{\text{I}}_{1}))$ is concave over $Q(g^{\text{I}}_{1}) \ge 0$, this problem is convex. By checking the first-order derivative of the objective function, the optimal solution to problem (\ref{eqn:25}) is derived as $\bar{Q}_{\mu}$ (irrespective of $g_1^{\text{I}}$). If $\mu \le \frac{\lambda_{0}\bar{\gamma}}{\lambda_{2}P_0}e^{-(\lambda_{0}\sigma^{2}_{0}+\lambda_{1}\sigma^{2}_{1})\bar{\gamma}/P_0}$, we have
\begin{align}\label{eqn:Q2}
\bar{Q}_{\mu} = \sqrt{\frac{\lambda_{2}P_0}{\lambda_{0}\bar{\gamma}\mu}e^{-(\lambda_{0}+\lambda_{1})\sigma^{2}\bar{\gamma}/P_0}} - \frac{\lambda_{2}P_0}{\lambda_{0}\bar{\gamma}},
\end{align}
with the achieved objective value given as
\begin{align}
v_\mu
=& \bigg(\sqrt{\frac{\mu\lambda_2 P_0}{\lambda_{0}\bar{\gamma}}} - \sqrt{e^{-(\lambda_{0}\sigma^{2}_{0}+\lambda_{1}\sigma^{2}_{1})\bar{\gamma}/P_0}}\bigg)^2 + \Phi(0).
\end{align}
Otherwise, if $\mu > \frac{\lambda_{0}\bar{\gamma}}{\lambda_{2}P_0}e^{-(\lambda_{0}\sigma^{2}_{0}+\lambda_{1}\sigma^{2}_{1})\bar{\gamma}/P_0}$, we have $\bar{Q}_{\mu} = 0$ and $v_\mu = \Phi(0)$.

By comparing $v_0(g_1^{\text{I}})$ versus $v_\mu$ for the two regimes, the optimal solution to problem (\ref{eqn:sub:pro}) under $g^{\text{I}}_{1} < \bar g_1$ is given as
\begin{align}\label{eqn:QQQ:2}
\tilde{Q}(g^{\text{I}}_{1}) =
&\left\{
\begin{array}{ll}
\bar{Q}_{\mu}, & {\text{if}} ~v_\mu >  v_0(g_1^{\text{I}})\\
0, & {\text{if}} ~v_\mu <  v_0(g_1^{\text{I}}),
\end{array}
\right.,\forall g^{\text{I}}_{1}<\bar g.
\end{align}

\subsubsection{Finding Optimal $\mu \ge 0$ to Solve (D2)}
Next, we solve (D2) to find the optima $\mu$, denoted by $\mu^\star$. It is easy to show that at the optimality of (P2), the average jamming power constraint must be tight. Therefore, the optimal $\mu^\star$ can be found by using the equation $\mathbb{E}_{g^{\text{I}}_{1}}(\tilde{Q}(g^{\text{I}}_{1})) = Q_{\text{ave}}$, with $\tilde{Q}(g^{\text{I}}_{1})$ given in \eqref{eqn:QQQ:1} and \eqref{eqn:QQQ:2}.

Under $\mu^\star$, the corresponding $\tilde{Q}(g^{\text{I}}_{1})$'s in \eqref{eqn:QQQ:2} become the optimal jamming power allocation ${Q}^\star(g^{\text{I}}_{1})$'s for (P2). After some simple manipulation, we can further show that \eqref{eqn:mu:star} must hold in order for the jamming power constraint to be tight. With \eqref{eqn:mu:star}, we can obtain \eqref{Q:opt} based on \eqref{eqn:QQQ:2}. Hence, this proposition is proved.

\ifCLASSOPTIONcaptionsoff
  \newpage
\fi

\bibliographystyle{IEEEtran}

\end{document}